\documentclass{article}
\usepackage[T1]{fontenc}
\usepackage{spconf,amsmath,amssymb,graphicx,booktabs}
\usepackage[nocompress]{cite}

\usepackage{hyperref}
\usepackage{subcaption}
\hypersetup{hidelinks,pdfauthor={Mu Jia, Junting Chen, Pooi-Yuen Kam},pdfsubject={Author discussion draft; verified experimental results}}
\renewcommand{\footnotesize}{\fontsize{9.1}{11.2}\selectfont}
\renewcommand{\small}{\fontsize{9.1}{11.2}\selectfont}
\title{RF-Florence: Unified RF Detection and\\Physical Grounding with a Foundation Model}
\name{Mu Jia$^{*\dagger\ddagger}$, Junting Chen$^{*\ddagger}$, and Pooi-Yuen Kam$^{\dagger\ddagger}$}
\address{$^{*}$School of Science and Engineering, $^{\dagger}$School of Artificial Intelligence,\\
$^{\ddagger}$Shenzhen Future Network of Intelligence Institute (FNii-Shenzhen),\\
The Chinese University of Hong Kong (Shenzhen), Guangdong 518172, China}
\begin{document}
\maketitle
\begin{abstract}
Reliable signal detection is essential for spectrum sharing and intelligent sensing in sixth-generation (6G) wireless networks. Wideband radio-frequency (RF) monitoring must locate emissions and identify those satisfying waveform-family and physical constraints. The latter task, physical grounding, requires query meaning to agree with measurable signal attributes. RF-Florence unifies three tasks---full-scene detection, physical grounding, and supplied-region recognition---through a shared pretrained foundation model. Calibrated spectrograms connect localization coordinates to time and frequency, while common event targets and explicit predicates align task outputs with requests. On synthetic mixtures with family--attribute combinations withheld from training, structured queries improve mean grounding F1 by 7.66 percentage points over natural language. Under the tested two-path channel, direct grounding with structured queries exceeds a You Only Look Once (YOLO) detector followed by rule-based filtering by 3.44 points in mean F1. These results show the value of aligning query expressions with the signal families and physical variables that define the requested event set.
\end{abstract}

\section{Introduction}
Signal detection and spectrum monitoring give 6G receivers the information they need to share spectrum and sense the radio environment~\cite{axell2012spectrum}. They support activity sensing~\cite{mazzieri2024microdoppler,lyons2025wisensenet}, integrated sensing and communication~\cite{zhang2021jcr,liu2020jointbeamforming}, unmanned aerial vehicle monitoring~\cite{liang2026uav}, and mobile edge reception~\cite{jia2026noncoherent}. Beyond locating emissions, receivers must identify which signals satisfy a sensing or communication request.

\begin{figure*}[!t]
\centering
\captionsetup[subfigure]{font={small,bf},labelfont=bf,skip=1pt}
\setlength{\fboxsep}{3pt}
\setlength{\fboxrule}{0.3pt}
\fbox{\begin{subfigure}{\dimexpr\textwidth-2\fboxsep-2\fboxrule\relax}
\caption{Calibrated RF observations and shared task outputs}
\includegraphics[width=\linewidth]{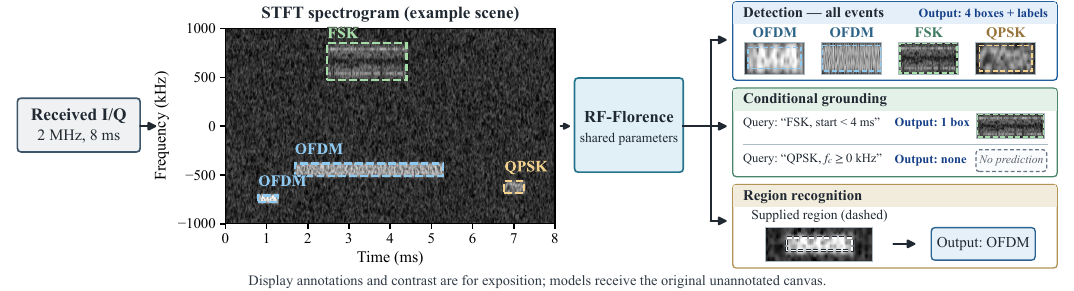}
\end{subfigure}}
\par\vspace{3pt}
\fbox{\begin{subfigure}{\dimexpr\textwidth-2\fboxsep-2\fboxrule\relax}
\caption{Query alignment with waveform families and physical conditions}
\includegraphics[width=\linewidth]{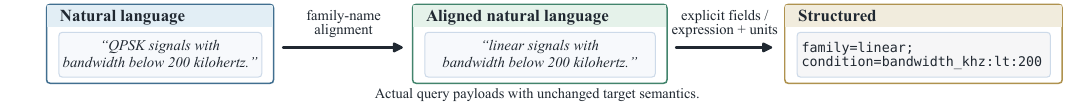}
\end{subfigure}}
\caption{RF-Florence framework. (a) One calibrated spectrogram supports multitask inference through shared model parameters; shown outputs are saved NL-model predictions on a development scene. (b) Family-name alignment and explicit physical conditions preserve query targets; QPSK instantiates \texttt{linear}. Models receive the unannotated canvas; crops and gamma correction (1.8) are display-only. Queries are abbreviated; $f_{\mathrm c}$ denotes center frequency. The common task token is omitted in (b), whose arrows compare query encodings.}
\label{fig:prompt_workflow}
\end{figure*}

Wideband monitoring may require all emission locations and families~\cite{vagollari2023wideband}, only emissions satisfying a physical condition, or the family of a supplied time--frequency region. We call the conditional localization task \emph{physical grounding}, such as locating narrowband emissions of a requested family. Noise uncertainty limits detection reliability~\cite{tandra2008snrwalls}; overlapping signals hinder time--frequency localization~\cite{zhao2024trtfl}, while multipath fading distorts features used for recognition~\cite{oshea2018overair}. Grounding further requires the family and physical attributes to jointly satisfy the request.

Classical spectrum sensing exploits compressive spectral estimation~\cite{polo2009compressive} or cyclostationary structure~\cite{cohen2017cyclostationary}. Learned spectrogram detectors, including YOLO models, extend this to event localization and recognition~\cite{boegner2022wbsig,vagollari2023wideband,mazouz2026multires}. Physical rules can efficiently filter detections, but cannot recover missed events; attribute errors can also change which events qualify. Wireless foundation models offer reusable representations for few-shot detection and spectral analysis~\cite{hao2025vslm,kim2026lwmspectro}, while instruction-based models broaden RF understanding~\cite{zou2026rfgpt,bara2026rfanalyzer}. Task-conditioned generation provides a common interface for heterogeneous outputs~\cite{xiao2024florence2}. For RF, this interface must also align coordinates, labels, and query expressions with the physical conditions defining target events.

Our goal is to unify RF detection, physical grounding, and region recognition over the same received observation. The central problem is to connect waveform identity and measurable conditions to the requested event set. Compositional, set-valued grounding has visual precedents~\cite{ding2026grex,liu2024finecops}; RF requests additionally depend on calibrated time and frequency. We study query encoding for unseen family--attribute combinations and direct grounding versus detect-then-filter processing under reception distortion. These questions guide the design of RF query interfaces and localization architectures.

RF-Florence therefore uses a common event representation for multitask RF analysis (Fig.~\ref{fig:prompt_workflow}). Calibrated spectrograms tie event geometry to physical quantities, while consistent family names and explicit predicates link queries to their targets. Each task returns a different part or subset of the same event description, making Florence-2's task-conditioned encoder--decoder a natural shared interface. Joint learning connects this interface to signal-derived supervision. Our contributions are:
\begin{itemize}
\setlength{\itemsep}{1pt}\setlength{\parskip}{0pt}
\brokenpenalty=10000 
\interlinepenalty=10000 
\item We develop a unified RF model using one shared encoder--decoder for full-scene detection, physical grounding, and supplied-region recognition.
\item We formulate grounding as predicates over waveform families and calibrated time--frequency attributes, connecting query semantics to common event targets.
\item We provide controlled evidence on query encoding, transfer to held-out family--attribute combinations, and direct grounding versus detect-then-filter processing under two-path reception.
\end{itemize}

\section{RF-Florence: Unified RF Detection and Grounding}
\label{sec:task}
\newcommand{\StoryTableRows}{%
\multicolumn{8}{@{}l}{\textit{RF-Florence variants}} \\
Natural language & 99.13 & 95.73 & 97.36 & 86.15 & 4.31 & 96.56 & 93.93 \\
\textbf{Structured} & \textbf{99.26} & 96.03 & \textbf{97.49} & 95.45 & 2.97 & \textbf{96.75} & 94.25 \\
Grounding-only & \NA & \NA & 97.42 & 88.19 & 2.06 & \NA & 93.65 \\
\addlinespace[3pt]
\multicolumn{8}{@{}l}{\textit{Reference systems}} \\
YOLO11s + executor & 98.07 & \textbf{96.55} & 97.34 & \textbf{96.14} & \textbf{0.99} & \NA & \textbf{29.32} \\
RT-DETRv2 + executor & 91.09 & 87.31 & 89.51 & 85.94 & 10.60 & \NA & 47.87 \\
Grounding DINO & 74.79 & 38.00 & 45.39 & 28.19 & 78.88 & \NA & 105.19 \\
MobileNetV3 & \NA & \NA & \NA & \NA & \NA & 96.53 & \NA \\
\addlinespace[2pt]
\bottomrule
}

\newcommand{\NA}{\multicolumn{1}{c}{--}}
\begin{table*}[!t]
\centering
\caption{RF detection, grounding, and region recognition results. Quality in percent; per-query latency in milliseconds}
\label{tab:prompt}
\small
\setlength{\tabcolsep}{3.0pt}
\begin{tabular*}{\textwidth}{@{\extracolsep{\fill}}lrrrrrrr@{}}
\toprule
& \multicolumn{2}{c}{Perception} & \multicolumn{3}{c}{Grounding} & Region & Latency\\
\cmidrule(lr){2-3}\cmidrule(lr){4-6}\cmidrule(lr){7-7}\cmidrule(lr){8-8}
System & Det. F1 $\uparrow$ & Rec. F1 $\uparrow$ & Seen F1 $\uparrow$ & Held-out F1 $\uparrow$ & Neg. FAR $\downarrow$ & Accuracy $\uparrow$ & (ms) $\downarrow$\\
\midrule
\StoryTableRows
\end{tabular*}
\par\vspace{2pt}\begin{minipage}{\textwidth}\footnotesize
Det./Rec.: detection/recognition. Neg. FAR: false-alarm rate for empty-target queries. Bold: best in column; ``--'': unavailable.
\end{minipage}
\end{table*}

\subsection{Physically calibrated RF representation}
RF-Florence connects event geometry, task outputs, and physical queries through a common representation and one shared encoder--decoder. The received in-phase and quadrature (I/Q) sequence $r[n]$ has sample index $n$, duration $T$, and sample rate $f_{\mathrm{s}}$. A fixed short-time Fourier transform (STFT) produces a calibrated spectrogram $S(r)$. Each emission has a normalized box $\mathbf b=(x_0,y_0,x_1,y_1)\in[0,1]^4$ giving top-left and bottom-right corners. Time increases rightward; frequency decreases downward. Calibration relates this geometry to duration, bandwidth, start time, and center frequency:
\begin{align}
 d(\mathbf b)&=T(x_1-x_0), & B(\mathbf b)&=f_{\mathrm{s}}(y_1-y_0),\label{eq:extent}\\
 t_0(\mathbf b)&=Tx_0, & f_{\mathrm c}(\mathbf b)&=\tfrac12 f_{\mathrm{s}}(1-y_0-y_1).\label{eq:position}
\end{align}
Calibration converts box extent into duration and bandwidth, and position into start time and center frequency. These physical quantities are independent of image size.

\subsection{Unified RF task interface}
For $K$ events indexed by $k$, the tasks share $(c_k,\mathbf b_k)$: family and support. Detection returns all events, $\mathcal E=\{(c_k,\mathbf b_k)\}_{k=1}^{K}$; region recognition estimates the family of a supplied support. For query $q$, grounding selects supports using a physical predicate $\phi_q$:
\begin{equation}
 \mathcal Y_q=\{\mathbf b_k:\phi_q(c_k,\mathbf b_k)=1\}.\label{eq:targets}
\end{equation}
Answers may contain zero, one, or several boxes. Signal metadata provide consistent targets for all tasks. Common event semantics motivates a shared encoder--decoder whose request determines which event information is returned.

\subsection{Physical queries and semantic alignment}
To align a request with its target set, write $q=(c_q,a_q,o_q,\tau_q)$ for a requested family, physical attribute, comparison operator, and threshold. For a candidate event $(c,\mathbf b)$ and a family-and-attribute request,
\begin{equation}
 \phi_q(c,\mathbf b)=\mathbf 1[c=c_q]\,\mathbf 1[g_{a_q}(\mathbf b)\;o_q\;\tau_q],\label{eq:predicate}
\end{equation}
where $g_{a_q}$ selects one of $d,B,t_0,f_{\mathrm c}$, $o_q\in\{<,\geq\}$, and $\mathbf 1[\cdot]$ is one when its condition holds and zero otherwise. An omitted family or attribute contributes a factor of one, accommodating family-only and attribute-only requests.

This rule composes family identity and attribute constraints, linking query terms to the labels and physical quantities defining the target set.

The supervised labels \texttt{ofdm}, \texttt{linear}, and \texttt{fsk} denote orthogonal frequency-division multiplexing (OFDM), linear modulation, and frequency-shift keying (FSK). The generator instantiates \texttt{linear} with single-carrier quadrature phase-shift keying (QPSK). Natural-language (NL) queries call this instance ``QPSK,'' whereas detection and region supervision use \texttt{linear}. Aligned NL replaces ``QPSK'' with \texttt{linear} to test consistency with the supervised family vocabulary. Structured queries additionally make family, attribute, operator, and threshold explicit. All encodings retain the same target sets (Fig.~\ref{fig:prompt_workflow}(b)); structure changes format and unit notation together.

\subsection{Shared foundation-model learning}
Coordinate tokens and waveform labels provide a common output vocabulary, so a task-conditioned decoder can express multiple tasks without separate prediction heads. RF-Florence uses encoder--decoder $F_\theta$ with shared parameters $\theta$ to generate $\hat y=F_\theta(S(r),u)$, where $\hat y$ is the output token sequence and $u$ encodes the task and any query or supplied region. Starting from pretrained Florence-2, $F_\theta$ is jointly fine-tuned on multitask examples using token-level cross-entropy.

\section{Experimental Design}
\label{sec:experiments}
\subsection{RF scenes and held-out compositions}
Synthetic scenes mix OFDM, QPSK, and FSK, with randomized power, placement, and noise. Overlap tests event separation; empty scenes test rejection. We use 3,584 training and 512 development records, separated by scene.

The split tests composition of familiar concepts: every family and attribute appears in training, but six conjunctions are withheld---OFDM long or wide, QPSK short or narrow, and FSK short or narrow. Grounding must combine these known primitives in untrained pairings. We use fixed thresholds of 800~$\mu$s for duration, 200~kHz for bandwidth, 4~ms for start time, and 0~kHz for center frequency.

\subsection{Model variants and reference systems}
Comparisons examine query encoding, task allocation, and inference route. Matched NL, aligned NL, and structured variants share initialization, seed, and update count. A grounding-only NL control allocates all updates to grounding. Repeated-training comparisons use three NL and three structured instances.

YOLO11s~\cite{jocher2024yolo11} and RT-DETRv2 with a ResNet-50 backbone~\cite{lv2024rtdetrv2} implement \emph{detect-then-filter} processing: exact predicates select detected events without a language parser. This route is labeled ``+ executor'' in the table and figure. Grounding DINO Tiny~\cite{liu2023groundingdino} and MobileNetV3 small~\cite{howard2019mobilenetv3} provide grounding and region references. All systems share underlying training records and use development selection.

\subsection{Paired nominal and two-path reception}
The \emph{unified-task set} evaluates all interfaces on 1,024 nominal scenes. The \emph{paired reception set} tests 512 new scenes under nominal and energy-normalized two-tap reception (4~$\mu$s, $-6$~dB echo). Paired observations share scene parameters and noise; supports include the echo tail. Both sets follow model and calibration freezing.

The echo changes spectral texture and can perturb estimated supports and their physical attributes. This links reception distortion to predicate evaluation, testing whether physical grounding remains useful under altered observations.

\subsection{Metrics and comparison scope}
One-to-one matching at intersection over union $0.5$ gives counts of true positives (TP), false positives (FP), and false negatives (FN). Their F1 score is $\mathrm{F1}=2\mathrm{TP}/(2\mathrm{TP}+\mathrm{FP}+\mathrm{FN})$. Detection matches geometry; recognition also requires family; grounding scores the requested target set. Region accuracy and empty-target false-alarm rate measure classification and rejection separately. Intervals use paired base-scene resampling, retaining the dependence among requests from the same observation.

Batch-one latency measures the fresh-request path from in-memory I/Q through frontend, inference, parsing, and filtering on an RTX~4090.

\setcounter{topnumber}{1}
\begin{table}[t]
\centering
\caption{Query-encoding gains: a matched single-run comparison}
\label{tab:query_effects}
\small
\setlength{\tabcolsep}{3pt}
\begin{tabular*}{\columnwidth}{@{\extracolsep{\fill}}lrr@{}}
\toprule
Query change & \shortstack{F1 (\%)\\Before $\to$ after} & \shortstack{Gain\\(pp)}\\
\midrule
NL $\to$ Aligned NL & $87.15 \to 90.59$ & $+3.44$\\
Aligned NL $\to$ Structured & $90.59 \to 96.58$ & $+5.99$\\
\midrule
NL $\to$ Structured (total) & $87.15 \to 96.58$ & $\mathbf{+9.43}$\\
\bottomrule
\end{tabular*}
\par\vspace{2pt}\begin{minipage}{\columnwidth}\footnotesize
NL: natural language; pp: percentage points.
\end{minipage}
\end{table}

\begin{figure}[t]
\centering
\includegraphics[width=\columnwidth]{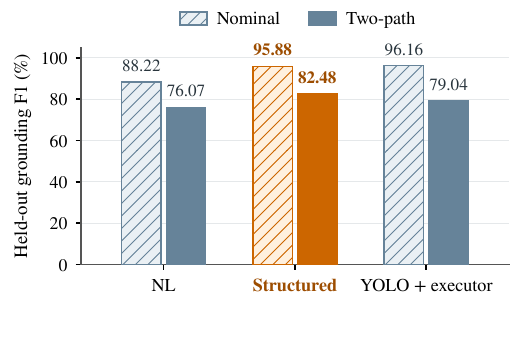}
\caption{Held-out grounding on 512 paired scenes, with 3,072 requests per condition. Hatched/solid bars show nominal/two-path F1. NL and Structured report three-run means; YOLO uses one model. Channel: 8-sample (4~$\mu$s), $-6$~dB echo.}
\label{fig:reception}
\end{figure}

\section{Results and Design Implications}
\label{sec:results}
\subsection{Shared parameters support multitask RF analysis}
One shared encoder--decoder supports multiple RF tasks with high accuracy (Table~\ref{tab:prompt}), using their common event representation. The specialized portfolio provides the same interfaces through detection, filtering, and classification.

The challenge is connecting requests to events: NL and structured variants have similar detection F1 but different held-out grounding performance. Allocating all updates to NL grounding also leaves a gap. These results motivate testing query encoding under matched supervision.

\subsection{Semantic alignment improves grounding}
Structured physical queries improve mean held-out grounding F1 by $7.66$ percentage points (pp) across three training runs, with 99\% paired interval $[6.19,9.15]$. The matched single-run ablation in Table~\ref{tab:query_effects} shows that family-name alignment provides part of the gain, with a further improvement from structured expression.

The gains extend beyond lexical consistency: exposing conditions in the physical variables defining target sets further improves grounding. This supports aligning both terminology and physical expression for compositional transfer, with format and units changed together.

\subsection{Physical grounding under two-path reception}
All systems degrade under two-path reception (Fig.~\ref{fig:reception}), but structured grounding exceeds YOLO filtering by $3.44$~pp. NL and structured results are three-run means; YOLO uses one model.

Direct grounding conditions localization on the request; filtering first commits to a catalogue. Its missed events cannot be recovered by rules, and label or support errors can alter predicate outcomes. This structural dependence motivates direct grounding under distorted reception. The measured advantage supports this design for the tested channel.

\subsection{Recall and processing cost}
As a secondary operating-point analysis, we sweep confidence thresholds on saved test predictions. At ten false-positive boxes per 100 requests, direct grounding attains higher two-path recall than detect-then-filter using either YOLO or the same RF-Florence model. Table~\ref{tab:prompt} also shows higher request latency than YOLO-based filtering, highlighting a tradeoff between grounding accuracy and processing cost.

\section{Conclusion}
RF-Florence unifies detection, physical grounding, and region recognition through a shared foundation-model interface. Calibrated geometry and common event targets connect these tasks to measurable RF quantities. Consistent naming and structured predicates improve held-out grounding, while direct grounding achieves higher F1 than detect-then-filter under the tested two-path channel. The central design insight is to align model queries with the waveform families and physical variables that define the requested signal set.

\clearpage
\begingroup
\raggedbottom
\let\rfOriginalBibliography\thebibliography
\renewcommand{\thebibliography}[1]{%
  \fontsize{9}{10.45}\selectfont
  \rfOriginalBibliography{#1}%
  \setlength{\itemsep}{4pt}%
  \setlength{\parsep}{0pt}%
  \setlength{\parskip}{0pt}%
}
\interlinepenalty=10000 
\bibliographystyle{IEEEbib-normalized}
\bibliography{references}
\endgroup
\end{document}